\documentstyle[eqsecnum,aps,epsfig,twocolumn]{revtex}

\begin{document}
\draft \preprint{}
\title{Exact numerical solution for a time-dependent fiber-bundle
model \\ with continuous damage }
\author{L. Moral}
\address{
Departamento de Matem\'atica Aplicada, Universidad de Zaragoza, \\
50009 Zaragoza, Spain.}
\author{J.B. G\'{o}mez}
\address{
Departamento de Ciencias de la Tierra, Universidad de Zaragoza, \\
50009 Zaragoza, Spain.}
\author{Y. Moreno}
\address{
The Abdus Salam International Centre for Theoretical Physics,\\
Condensed Matter Group, P.O. Box 586, Trieste, I-34014, Italy. }
\author{A.F. Pacheco}
\address{
Departamento de F\'{\i}sica Te\'orica, Universidad de Zaragoza, \\
50009 Zaragoza, Spain.}

\date{\today}
\maketitle
\begin{abstract}
A time-dependent global fiber-bundle model of fracture with
continuous damage was recently formulated in terms of an
autonomous differential system and numerically solved by applying a
discrete probabilistic method. In this paper we provide a method
to obtain the exact numerical solution for this problem. It is
based on the introduction of successive integrating parameters
which permits a robust inversion of the numerical integrations
appearing in the problem.
\end{abstract}

\narrowtext

\section{Introduction}
\label{sec:intro}

Fracture in disordered media has for many years attracted much
scientific and industrial interest
\cite{h90,cha,garcia97,maes98,petri94,zape97,prlus00}. An
important class of models of material failure is the fiber-bundle
models (FBM) which have been extensively studied during the past
decades \cite{prlus00,col57,pho83,new94,new95,vaz99}. These models
consist of a set of parallel fibers having statistically
distributed strengths. The sample is loaded parallel to the fiber
direction, and a fiber fails if the load acting on it exceeds a
strength threshold value. When a fiber fails, its load is
transferred to other surviving fibers in the bundle, according to
a specific transfer rule. Among the possible options of load
transfer, one simplification that makes the problem analytically
tractable is the assumption of equal load sharing (ELS), or global
load transfer, which means that after each fiber breaks, its
stress is equally distributed among the intact fibers. Until very
recently, the failure rule applied in standard FBM was
discontinuous and irreversible, i.e., when the local load exceeds
the failure threshold of a fiber, the fiber is removed from the
calculation and is never restored. Recently, a novel continuous
damage law was incorporated into these models \cite{nat97,kun00}.
Thus, when the strength threshold of a fiber is exceeded, it
yields, and the elastic modulus of the fiber is reduced by a
factor $a$ ($0<a<1$). Multiple yields of a given fiber are
allowed, up to a maximum of $n$ yielding events per fiber, where
$n$ is a small integer number which can be different for each
fiber. This generalization of the standard FBM is suitable to
describe a variety of elasto-plastic constitutive behaviors
\cite{na95,evans94,kanada99}.

The standard FBMs simulate the failure of a system at the {\em
microscopic} level. Each fiber breakage can be mapped onto a new
microcrack (with a typical size of a few $\mu$m), or onto the
extension of a previous microcrack. On the other hand, the
continuous damage FBMs simulate failure at a {\em mesoscopic}
level. Now, each fiber in the model can be viewed as a small
volume of the material. The term ``small'' depends on the size of
the heterogeneities, but can be of the order of one millimeter for
rocks. In each of these representative elementary volumes (REVs)
in which the total volume can be divided, there are many potential
sites for crack nucleation and growth, and the addition of each
new crack will change continuously the elastic properties of the
REV until its final failure when the accumulated damage surpasses
a threshold. This threshold is identified in our model with the
parameter $n$. Another important parameter in the model, the
stiffness reduction factor, $a$, controls the amount of weakening
that each yielding event introduces in a REV. The value $a=1$
means  no weakening, so that the elastic modulus of the REV
remains the same irrespective of the number of yieldings, a rather
unphysical situation. At the other extreme, the value $a=0$ means
complete weakening after the first yield event. Thus, $0 < a < 1$
is the physically meaningful range for the stiffness reduction
factor. In all the results given in the following sections, we
have assumed that the initial elastic module of all the REVs is
unity and that $n$ is the same for all the fibers. The randomness
is incorporated in the REVs' liftimes, not in the elastic moduli.

FBMs come in two settings, static and time-dependent or dynamic,
and both of them have been applied to the standard and continuous
damage settings \cite{nat97,kun00,moral01,kun01}. The static
version of FBM simulates the failure of materials by quasistatic
loading. Drawing an analogy with what is carried out in a
deformation experiment in the laboratory, a static FBM simulates a
uniaxial or triaxial, compressive or tensile, deformation test
where the duration of the test is measured in seconds or minutes.
In these models, the stress on each fiber is the independent
variable and the strength of each element is the distributed
random variable. On the other hand, the dynamic FBM simulates
failure by creep rupture, static fatigue, or delayed rupture,
i.e., a (usually) constant load is imposed on the system and the
elements break because of fatigue after a period of time. The time
elapsed until the system collapses is the lifetime of the bundle.
Time acts as an independent variable, and the initial lifetime of
each element, for a prescribed initial stress, is the independent
identically distributed random quantity. Again, we can draw a
clear analogy with a particular type of deformation experiments in
the laboratory, the so-called creep experiments, where a
heterogeneous material (rock, concrete, composite, ceramic alloy)
is subjected to a constant or cyclic load, breaking after a period
of time. The duration of these tests depends on the load imposed
on the material and, more exactly, on the load compared to the
short-term strength of the material ({\em i.e.}, the load that
causes the ``instantaneous'' failure of the same material in a
fast uniaxial experiment). This load is usually expressed as a
percentage of the short-term strength and the duration of the
experiments is critically dependent on it. For rock, say, a sample
will fail by creep after a few hours when subjected to a load 80\%
of the short-term strength, after a few weeks for a load 70\% of
the short-term strength, and after a few months or even years for
lower working loads. The mechanism behind creep failure is {\em
subcritical crack growth}, {\em i.e.}, the slow extension of
microcracks with lengths smaller than the critical crack length
for instantaneous failure. Subcritical crack growth is due to a
variety of processes operating near crack tips, the most important
of them being {\em stress corrosion}, a chemical interaction
between the crack tip and the environmental species, notably
water, filling the microcracks that provokes the hydrolytic
weakening of the atomic bonds of the material in the crack tip,
where stress concentrations are highest. The crack propagation
velocity is extremely sensitive to the applied load, suggesting
exponential or power-law velocity functions with large
coefficients or exponents.

Indeed, in the dynamic FBMs the most widely used breaking rate
function is the power law \cite{new94,new95,vaz99}, in which
elements break at a rate proportional to a power of their stress,
$\sigma^{\rho}$, where the exponent $\rho$ is an integer called
the stress corrosion exponent, for obvious reasons. This type of
breaking rate will be assumed here and is another parameter of the
model.

Our generalization of the dynamic global FBM \cite{moral01} was
restricted to the global transfer modality, and there we assumed
that the size of the bundle, $N$, was very large. This enabled us
to formulate the evolution of the system in terms of continuous
differential equations. This type of equation, similar to those
appearing in radioactivity, was first used by Coleman
\cite{col57}, and later in \cite{new95}. In \cite{moral01} we
supposed an ELS bundle formed by $N$ fibers which breaks because
of stress corrosion under the action of an external constant load
$F=N\cdot \sigma_0$, with $\sigma_0=1$. The breaking rate of the
fibers, $\Gamma$, is assumed to be of the power-law type,
$\Gamma=\sigma^{\rho}$, $f$ denotes the strain of the bundle and
$Y=1$ represents the initial stiffness of the individual fibers.
The original dynamic FBM was generalized by allowing one fiber to
fail more than once, and thus we define the integer $n$ as the
maximum value of failures allowed per fiber. Besides, as mentioned
before, the parameter $a$ ($<1$) represents the factor of
reduction in the stiffness of the fibers when they fail. As up to
$n$ partial yielding events are permitted per fiber, at any one
time the population of fibers will be sorted in $n+2$ lists. Thus
$ N=N_0+N_1+\dots+N_n+N^{\prime}, \label{eq3} $ where $N_i$
($i=0,\dots,n$) denotes the number of elements that have failed
$i$ times. $N^{\prime}$ denotes the number of elements that have
failed $n+1$ times and therefore are inactive (i.e., they no
longer support any load anymore). At $t=0$, the $N$ elements of
the bundle form the list $0$, $N_0=N$, and at $t=T$,
$N^{\prime}=N$. The specification, at a given time $t$, of the
value of $N_i$, for $i=0,1,\dots,n$, provides the state of the
system. In our continuous formulation the $N_i$ are real positive
numbers lower than $N$.

As the external load $F=N$ is supported by the present active
fibers, we have $N=f\cdot(N_0+a N_1+a^{2}N_2+\dots+a^{n}N_n),
\label{eq1.2} $ and hence
\begin{equation}
f=N/(N_0+aN_1+a^{2}N_2+\dots+a^{n}N_n). \label{eq1.1}
\end{equation}

The time evolution equations are \cite{moral01}:
\begin{eqnarray}
 \frac{dN_0}{dt}&=&f^{\rho}(-N_0), \nonumber \\
 \frac{dN_1}{dt}&=&f^{\rho}(N_0-\kappa N_1), \nonumber \\
 \frac{dN_2}{dt}&=&f^{\rho}\kappa(N_1-\kappa N_2), \label{eq1.3} \\
 \vdots & &\qquad \qquad \vdots \nonumber \\
 \frac{dN_n}{dt}&=&f^{\rho}\kappa^{n-1}(N_{n-1}-\kappa N_n), \nonumber
\end{eqnarray}
where the ubiquitous constant factor $\kappa$ represents
$\kappa=a^{\rho}$. This is an autonomous differential system. Its
solution must fulfill the initial condition
\begin{eqnarray}
N_0(t=0)&=&N \nonumber \\ N_j(t=0)&=&0, \quad j\ne 0.
\label{eq1.4}
\end{eqnarray}

An alternative way of introducing a time-dependent rheological
response in FBM is that of Cruz-Hidalgo {\em et. al.}
\cite{kun01}. These authors incorporate a viscoelastic
constitutive behaviour in their model through the mapping of each
fiber to a Kelvin-Voigt element. They express the time evolution
of the strain in {\em each} fiber by way of a differential
equation. In their model, fibers break irreversibly when they
surpass a statistically distributed strain threshold, whereas in
our model multiple failures (yields) of a fiber are allowed, the
variable which is statistically distributed is the lifetime of the
fibers, and there is no explicit threshold dynamics. This
different formulation implies that we can formulate the evolution
of the system in terms of coupled differential equations, while
the authors in ref. \cite{kun01} have necessarily to use Monte
Carlo simulations due to the lack of a global differential
equation for the system.

In reference \cite{moral01}, Eqs.\ (\ref{eq1.3}) were solved by
applying a numerical probabilistic method. The purpose of this
paper is to present an exact numerical method that solves Eq.\
(\ref{eq1.3}), fulfilling the initial conditions (\ref{eq1.4}).
This method is explained in Section \ref{sec:exact}. In Section
\ref{sec:results} we present a discussion of the method and of the
results. The reader will find a longer discussion of the physical
results in Ref. \cite{moral01}. This paper concentrates on the
solution method.

\section{EXACT NUMERICAL METHOD}
\label{sec:exact}

To simplify the notation, we first normalize the variables
\begin{equation}
x_i=\frac{N_i}{N}, \quad i=0,1,\cdots,n. \label{eq2.1}
\end{equation}
In terms of the $x_i$, the differential system to be solved is

\begin{eqnarray}
\dot{x}_0 &=& -f^\rho x_0 \nonumber \\ \dot{x}_j &=& f^\rho
\kappa^{j-1}(x_{j-1} - \kappa x_j) \label{eq2.2} \\ x_0(0)&=&1,
\quad x_j(0)=0, \quad j=1,2,\cdots,n.\nonumber
\end{eqnarray}
A dot on a variable means derivation with respect to time, and $f$
and $\kappa$ are the same objects as in Section \ref{sec:intro}:

\begin{equation}
1/f = \sum_{i=0}^{n} a^i x_i. \label{eq2.3}
\end{equation}

The system (\ref{eq2.2}) admits a reduction of degrees of freedom
by eliminating $t$ from the last $n$ equations and by integrating
with respect to $x_0$:

\begin{eqnarray}
\dot{x}_0 &=& -f^\rho x_0, \nonumber \\ \frac{dx_j}{dx_0} &=&
\frac{ \kappa^{j-1}(\kappa x_{j} - x_{j-1})}{x_0}. \label{eq2.4}
\end{eqnarray}
From Eq.\ (\ref{eq2.4}) we obtain

\begin{equation}
x_j = \sum_{l=0}^{i} b_l^{(i)} x_0^{\kappa^l}, \quad
i=0,1,\cdots,n, \label{eq2.5}
\end{equation}
with
\begin{eqnarray}
b_0^{(0)} &=& 1 \nonumber \\ b_l^{(j)} &=&
\frac{b_l^{(j-1)}\kappa^{(j-1)}}{\kappa^j - \kappa^l} \nonumber \\
b_j^{(j)} &=& - \sum_{l=0}^{j-1} b_l^{(j)}, \quad j=1,2,\cdots,n.
\label{eq2.6}
\end{eqnarray}
In consequence,

\begin{eqnarray}
f &=& \frac{1}{\sum_{i=0}^{n} a^i x_i} = \frac{1}{\sum_{i=0}^{n}
\alpha_i x_0^{\kappa^i}} =: f_0(x_0) \nonumber \\ \alpha_i &=&
\sum_{l=i}^{n} a^l b_l^{i}. \label{eq2.7}
\end{eqnarray}
Then, the first equation in (\ref{eq2.4}) provides the relation
$x_0$ versus $t$
\begin{equation}
t = \int_{x_0}^{1} \frac{dx_0}{\left[ f_0(x_0) \right]^{\rho} x_0}
= \int_{x_0}^{1} \frac{\left( \sum_{i=0}^{n} \alpha_i
x_0^{\kappa^i} \right)^\rho}{x_0}dx_0 \label{eq2.8}
\end{equation}
which, in principle, solves the problem because it relates $t$ to
$x_0$ and hence to any other $x_j$. However, the integral
(\ref{eq2.8}) is, in general, improper for $x_0 \rightarrow 0$
because the integrand is $O(x_0^{\rho\kappa^n-1})$ and therefore
the numerical relation $t$ vs.\ $x_0$ is problematic.
Specifically:

a) If $\rho\kappa^n-1 \le 0$ this integral is proper,

b) if $\rho\kappa^n-1 > 0$ the integral is improper. \\ Due to
the fact that the convergence occurs iff $\rho\kappa^n-1
> -1$, Eq.\ (\ref{eq2.8}) is always convergent, because in our
model of fracture $\rho\kappa^n > 0$.

Let $\epsilon \in (0,1)$; due to the fact that $x_0$ decays from 1
to 0, there exists a time value $t_0 > 0$ such that $x_0(t_0) =
\epsilon$. If (\ref{eq2.8}) is improper, we perform the following
change of parameter: $x_0 \equiv y_0 \rightarrow y_1$, such that

\begin{eqnarray}
\dot{x}_0 &=& -f^\rho x_0 \nonumber \\ \dot{y}_1 &=& -\kappa
f^\rho y_1 \nonumber \\ \dot{x}_j &=& f^\rho \kappa^{j-1} \left[
x_{j-1} - \kappa x_j \right]; \quad j=1,2,\cdots,n, \label{eq2.9}
\end{eqnarray}
with $x_0(t_0)=\epsilon$, $y_1(t_0)=1$, and $t>t_0$. From here

\begin{equation}
\frac{dx_0}{dy_1} = \frac{\kappa x_0}{y_1} \Rightarrow x_0 = c_1
y_1^{1/\kappa}, \quad c_1 = \left( x_0(t_0) \right) = \epsilon.
\end{equation}
Hence
\begin{equation}
x_j = \sum_{l=0}^{j} b_l^{(j)} \left( \epsilon y_1^{1/\kappa}
\right)^{\kappa^l} = \sum_{l=0}^{j} \beta_l^{(j,1)}
y_1^{\kappa^{l-1}}, \label{eq2.11}
\end{equation}

\begin{equation}
f=:f_1(y_1) = \frac{1}{\sum_{i=0}^{n} \alpha_i \sum_{l=0}^{i}
\beta_l^{(i,1)} y_1^{\kappa^{l-1}}} = \frac{1}{\sum_{i=0}^{n}
\alpha_i^{(1)} y_1^{\kappa^{i-1}}}
\end{equation}
with $\beta_l^{(i,1)} = b_l^{(i)}\epsilon^{\kappa^l}$,
$\alpha_i^{(1)} = \sum_{l=i}^{n} \alpha_l \beta_i^{(i,1)}$
$(i=0,1,\cdots,n)$.

In these circumstances (\ref{eq2.11}) and the equation
\begin{equation}
t-t_0 = \int_{y_1}^{1} \frac{dy_1}{\left( f_1(y_1)\right)^\rho
y_1} = \int_{y_1}^{1} \frac{\left( \sum_{i=0}^{n} \alpha_i^{(1)}
y_1^{\kappa^{i-1}} \right)^\rho}{y_1}dy_1 \label{eq2.12}
\end{equation}
describe $t$, $x_0, \cdots, x_n$ in terms of the $y_1$ parameter,
for $t \ge t_0$.

As the integrand of (\ref{eq2.12}) is
$O(y_1^{\rho\kappa^{n-1}-1})$, then

a) If $\rho\kappa^{n-1}-1 \ge 0$ (\ref{eq2.12}) is a proper
integral,

b) if $\rho\kappa^{n-1}-1 < 0$ (\ref{eq2.12}) is an improper
integral, but (\ref{eq2.12}) is always convergent.

Now, as $y_1$ decays to zero, there exists a time instant $t_1 >
t_0$ such that $y_1(t_1) = \epsilon$. And by considering the
change of parameter $y_1 \rightarrow y_2$ given by the conditions

\begin{eqnarray}
\dot{x}_0 &=& -f^\rho x_0 \nonumber \\ \dot{y}_1 &=& -\kappa
f^\rho y_1 \nonumber \\ \dot{y}_2 &=& -\kappa^2 f^\rho y_2
\nonumber
\\ \dot{x}_j &=& f^\rho \kappa^{j-1} \left[ x_{j-1} - \kappa x_j
\right]; \quad j=1,2,\cdots,n, \label{eq2.13}
\end{eqnarray}
with $y_2(t_2)=\epsilon$, $y_2(t_1)=1$, and $t>t_1$, we have

\begin{equation}
\frac{dy_1}{dy_2} = \frac{\kappa y_1}{y_2} \Rightarrow y_1 = c_2
y_2^{1/\kappa}, \quad c_2 = \left( y_1(t_1) \right) = \epsilon,
\end{equation}
and hence
\begin{equation}
x_j = \sum_{l=0}^{j} b_l^{(j,1)} \left( \epsilon y_2^{1/\kappa}
\right)^{\kappa^{l-1}} = \sum_{l=0}^{j} \beta_l^{(j,2)}
y_2^{\kappa^{l-2}}, \label{eq2.14}
\end{equation}

\begin{equation}
f=:f_2(y_2) =  \frac{1}{\sum_{i=0}^{n} \alpha_i^{(2)}
y_2^{\kappa^{i-2}}}
\end{equation}
with identical meaning as before for $\beta_l^{(j,2)}$ and
$\alpha_j^{(2)}$. Then, (\ref{eq2.14}) and
\begin{equation}
t-t_1 = \int_{y_2}^{1} \frac{dy_2}{\left( f_2(y_2)\right)^\rho
y_2} = \int_{y_2}^{1} \frac{\left( \sum_{i=0}^{n} \alpha_i^{(2)}
y_2^{\kappa^{i-2}} \right)^\rho}{y_2}dy_2 \label{eq2.15}
\end{equation}
describe $t$, $x_0, \cdots, x_n$ in terms of $y_2$, for $t \ge
t_1$. Besides, as the integrand of (\ref{eq2.15}) is
$O(y_2^{\rho\kappa^{n-2}-1})$, then

a) if $\rho\kappa^{n-2}-1 \ge 0$ (\ref{eq2.15}) is a proper
integral,

b) if $\rho\kappa^{n-2}-1 < 0$ (\ref{eq2.15}) is an improper
integral, but always convergent.

The process followed so far is generalized in the way expressed in
Table \ref{tbl1} where in the end
\begin{equation}
f=:f_n(y_n) =  \frac{1}{\sum_{i=0}^{n} \alpha_i^{(n)}
y_n^{\kappa^{i-n}}}, \label{eq2.16}
\end{equation}
and therefore
\begin{equation}
t-t_{n-1} = \int_{y_n}^{1} \frac{dy_n}{\left( f_n(y_n)\right)^\rho
y_n} = \int_{y_n}^{1} \frac{\left( \sum_{i=0}^{n} \alpha_i^{(n)}
y_n^{\kappa^{i-n}} \right)^\rho}{y_n}dy_n, \label{eq2.17}
\end{equation}
whose integrand is $O(y_n^{\rho-1})$, that is, integral
(\ref{eq2.17}) is always proper.

\section{RESULTS AND CONCLUSIONS} \label{sec:results}
 The simple formalism written in Section
\ref{sec:exact} can be expressed, for example, in a brief program
of MATHEMATICA and its results graphically appreciated. We omit
here the program but it can be provided on request. By fixing the
constants at the following values: $n=3$, $a=0.6$, $\rho = 2$, and
$\epsilon = 0.1$, in Fig.\ 1 the value of the working parameters
$y_i$ are represented vs.\ time. Note that their range of
definition is from 1 to $\epsilon$, except for $y_3$ which ends at
0 for $t_3=T$, i.e., the actual lifetime of the bundle.

In Fig.\ 2 we again show the evolution of the working parameters
and also the evolution of the four lists $x_i$ of elements in the
problem.

The strategy developed in Section \ref{sec:exact} can be
summarized in a few sentences. First, let us observe Fig.\
\ref{fig2} to appreciate the time evolution of the different
lists: while $x_0$ monotonously declines from 1 at $t=0$ to 0 at
$t=T$, the lists $x_j$, $j=1,2,3$ start from 0 at $t=0$, rise to a
maximum and then monotonously decline to 0 at $t=T$ (strictly
speaking, all the lists vanish at the same time). The last list
$j=n$ is special in the sense that it is the only one that tends
to 0 with an infinite slope when $t$ tends to $T$.

The analytical resolution of Eq.\ (\ref{eq2.2}) is impossible
because of the nonlinearity introduced by the $f^\rho$ factors.
This source of complexity is partly overcome after having
recognized the partial reduction of degrees of freedom expressed
in (\ref{eq2.4}). This partial reduction leads to the relation
between $x_j$, $j=1,2,\cdots,n$ and $x_0$, hence from
(\ref{eq2.8}) one has solved in principle the time evolution of
$x_0$, and of the rest of $x_j$. But, in (\ref{eq2.8}) one also
recognizes that this integral is improper. This is the {\em real
problem} we face for the numerical inversion $t \leftrightarrow
x_0$ in the region where $x_0$ is very small. In intuitive terms,
this shows in Fig.\ \ref{fig2} because beyond a certain time,
$x_0$ is no longer significant and its relation with $t$ becomes
``delicate''. Therefore we have used $x_0=y_0$ as a good
integration parameter only up to $t=t_0$. Beyond this point we
successively introduced other ``artificial parameters''
$y_1,y_2,\cdots,y_n$ which in the corresponding time interval play
the role performed by $x_0$ from 0 to $t_0$. Using these
parameters, we are able to robustly relate all the variables $x_i$
to $t$ in the whole interval from 0 to $T$.

At the end of the process, the last integral is always proper,
which allows a robust numerical inversion in the vicinity of
$t=T$. Intuitively, this is clear in Fig.\ \ref{fig2} where we
appreciate the abrupt fall-off of $x_3$.

In the comments written in Section \ref{sec:exact} after Eqs.\
(\ref{eq2.8}), (\ref{eq2.12}), (\ref{eq2.15}), and (\ref{eq2.17})
regarding the nature of those integrands, we noted that in general
they behave as $O(y_i^{\rho\kappa^{n-i}-1})$. This implies that
the condition
\begin{equation}
\rho\kappa^{n-i}-1 \ge 0 \label{eq3.1}
\end{equation}
tells us the value of $i=i_c$,
\begin{equation}
i_c \ge n - \frac{\ln \rho}{|\ln \kappa|} \label{eq3.2},
\end{equation}
such that, for $i \ge i_c$, the respective integral is proper and
there is no need to introduce more artificial integrating
parameters.

The reader should note that the $\epsilon$ introduced in the
method is not a limiting factor of precision, but merely sets the
temporal ranges of the various integrating parameters $y_i$. In
our procedure, the only source of inaccuracy is the precision of
MATHEMATICA, used for the numerical inversion of the integrals.

As a final conclusion we would say that the exact numerical method
presented in this paper to solve this fiber-bundle problem does
not predict any new qualitative result with respect to what was
obtained using the approximate method of Ref. \cite{moral01}.
Therefore, no new physical conclusions can be drawn from here.

The use of this exact strategy in other scientific problems that
are cast as an autonomous differential system will be considered
in the next future. In this respect, clear candidates are some
ecological problems and models of infection spreading
\cite{rama81,ande92}.

\acknowledgments

This work was supported in part by the Spanish DGICYT Project
PB98-1594.

\begin{figure}[t]
\begin{center}
\epsfig{file=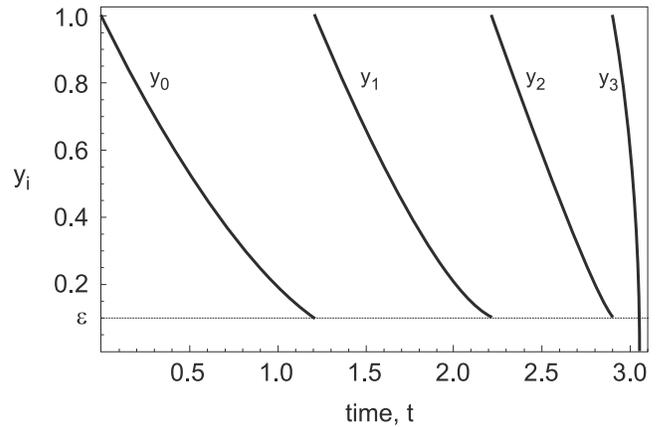,width=8.5cm,angle=0,clip=1}
\end{center}
\caption{Time evolution of the four integrating parameters $y_0$,
$ y_1$, $y_2$ and $y_3$ for a system with $n=3$, $a=0.6$ and
$\rho=2$. Note that their range of definition is from 1 to
$\epsilon$, except for $y_3$, which goes from 1 to 0.}
\label{fig1}
\end{figure}

\begin{figure}[t]
\begin{center}
\epsfig{file=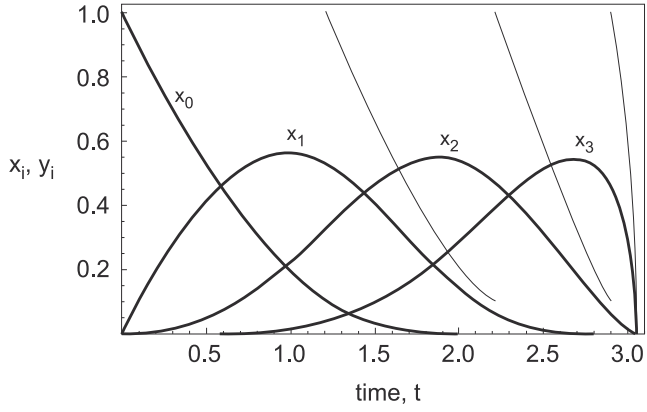,width=8.5cm,angle=0,clip=1}
\end{center}
\caption{Time evolution for the four integrating parameters and
the four variables $x_0$, $x_1$, $x_2$ and $x_3$ for a system with
the same parameters as for Fig.\ 1.} \label{fig2}
\end{figure}

\widetext
\begin{table}
\caption{General terms in the procedure.}
\begin{tabular}{ccc}
 Time interval   & Condition &  Parameter\\ \tableline
 $[0,t_0]$       &                             & $y_0=x_0$  \\
 $[t_0,t_1]$     & $y_0(t_0)=\epsilon$         &$y_1$ such that $\dot{y}_1 = -\kappa f^\rho y_1$; $y_1(t_0)=1$\\
 $[t_1,t_2]$     & $y_1(t_1)=\epsilon$         &$y_2$ such that $\dot{y}_2 = -\kappa^2 f^\rho y_2$; $y_2(t_1)=1$\\
 $\vdots$        & $\vdots$                    &$\vdots$\\
 $[t_{n-1},t_n]$ & $y_{n-1}(t_{n-1})=\epsilon$ &$y_n$ such that $\dot{y}_n = -\kappa^n f^\rho y_n$; $y_n(t_{n-1})=1$\\
 \end{tabular}
 \label{tbl1}
 \end{table}

\narrowtext

\end{document}